**Dr. Petar Radanliev**
Parks Road,
Oxford OX1 3PJ
United Kingdom
Email: petar.radanliev@cs.ox.ac.uk
BA Hons., MSc., Ph.D. Post-Doctorate

# Red Teaming Quantum-Resistant Cryptographic Standards: A Penetration Testing Framework Integrating AI and Quantum Security

*Abbreviated running title*: Red Teaming Quantum AI

Dr Petar Radanliev[1,2]

[1] Department of Computer Sciences, University of Oxford, Wolfson Building, Parks Rd, Oxford OX1 3QG, Email: petar.radanliev@cs.ox.ac.uk

[2] The Alan Turing Institute, British Library, 96 Euston Rd., London NW1 2DB

## Risk Scenarios

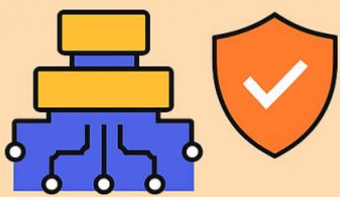

**Scenario One: Forward-Looking Quantum Threats**

A quantum computer forges digital signatures and decrypts current secure communications

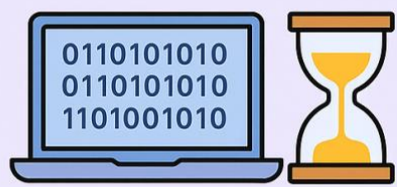

**Scenario Two: Retrospective Decryption and Historical Tampering ('Time-Travel' Attacks)**

Enables decryption of stored data, allowing manipulation of historical records

Dr. Petar Radanliev
Parks Road,
Oxford OX1 3PJ
United Kingdom
Email: petar.radanliev@cs.ox.ac.uk
BA Hons., MSc., Ph.D. Post-Doctorate

**Abstract**: Quantum computing and artificial intelligence are transforming the cybersecurity landscape, dictating a reassessment of cryptographic resilience in the face of emerging threats. This study presents a structured approach to evaluating vulnerabilities within quantum cryptographic protocols, focusing on the BB84 quantum key distribution method and National Institute of Standards and Technology (NIST) approved quantum-resistant algorithms. NIST is recognised globally as the leading institution in this domain, and the algorithms are selected though a decade-long process. By integrating AI-driven red teaming, automated penetration testing, and real-time anomaly detection, the research develops a robust framework for assessing and mitigating security risks in quantum networks. The findings demonstrate that AI can be effectively used to simulate adversarial attacks, probe weaknesses in cryptographic implementations, and refine security mechanisms through iterative feedback. The use of automated exploit simulations and protocol fuzzing provides a scalable means of identifying latent vulnerabilities, while adversarial machine learning techniques highlight novel attack surfaces within AI-enhanced cryptographic processes. This study offers a comprehensive methodology for strengthening quantum security and provides a foundation for integrating AI-driven cybersecurity practices into the evolving quantum landscape.

# 1. Introduction

The increasing integration of quantum computing and artificial intelligence (AI) is reshaping the cybersecurity landscape, forcing a reconsideration of traditional cryptographic frameworks. As quantum technologies progress, existing security paradigms must be evaluated for their resilience against AI-driven threats. This study focuses on the interaction between AI, particularly Natural Language Processing (NLP) models, and quantum security protocols, with a primary emphasis on the BB84 quantum key distribution (QKD) method and NIST-approved quantum-resistant algorithms. By using computational techniques in Python and C++, this research investigates the vulnerabilities within quantum cryptographic systems and explores how AI can enhance and undermine their security. The integration of machine learning techniques into cybersecurity has proven effective in enhancing threat

Dr. Petar Radanliev
Parks Road,
Oxford OX1 3PJ
United Kingdom
Email: petar.radanliev@cs.ox.ac.uk
BA Hons., MSc., Ph.D. Post-Doctorate

detection and response strategies, offering valuable insights into potential vulnerabilities within cryptographic systems [1].

This study adopts a penetration testing approach to analysing the security of quantum cryptographic protocols through adversarial simulation, commonly referred to as "red teaming." AI-driven adversarial testing allows for a systematic examination of potential weaknesses in quantum key exchange mechanisms, revealing attack vectors that conventional cryptographic assessments may overlook. The study extends beyond theoretical considerations by implementing real-world security testing methodologies, ensuring that quantum cryptographic frameworks are assessed under practical adversarial conditions.

## 2. Methodological Approach

This study adopts a multi-phase methodological framework designed to systematically evaluate the resilience of quantum cryptographic protocols, specifically the BB84 quantum key distribution (QKD) method and NIST-endorsed post-quantum cryptographic algorithms, against adversarial threats simulated through AI-driven red teaming. The methodological design integrates principles from cybersecurity testing, machine learning, and quantum cryptography, enabling a comprehensive and repeatable evaluation pipeline.

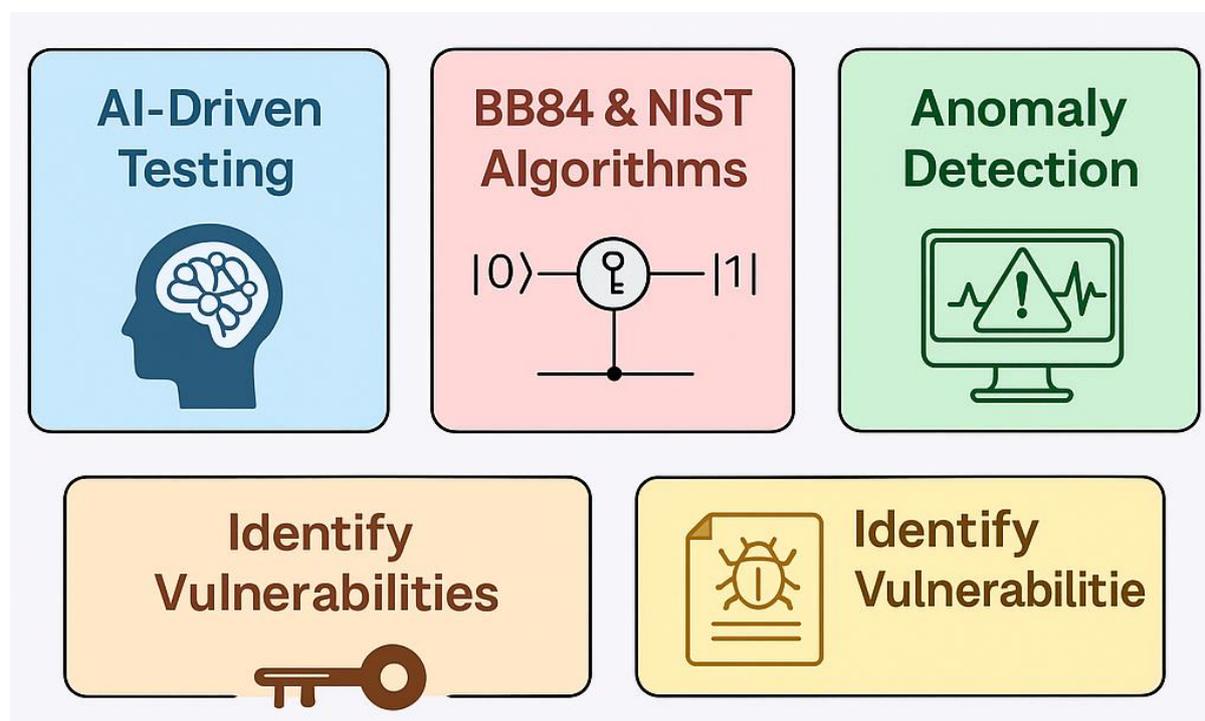

*Figure 1: Red Teaming Quantum-Resistant Cryptographic Standards: Framework integrating artificial intelligence and quantum security*

The research commences with a critical review of the literature on quantum cryptography, post-quantum cryptographic standards, and adversarial machine learning. This informs the construction of threat models aligned with realistic attack scenarios in both classical and quantum computing environments. A risk-informed

**Dr. Petar Radanliev**
Parks Road,
Oxford OX1 3PJ
United Kingdom
Email: petar.radanliev@cs.ox.ac.uk
BA Hons., MSc., Ph.D. Post-Doctorate

model selection process guides the identification of candidate cryptographic protocols for testing, prioritising those standardised by NIST and those susceptible to quantum-accelerated cryptanalysis.

The experimental phase employs an AI-enhanced penetration testing framework, structured around seven discrete stages: (1) threat modelling, (2) environment setup, (3) model training, (4) red teaming simulation, (5) anomaly detection, (6) reverse engineering and forensic analysis, and (7) remediation and refinement. Python is utilised for scripting, automation, and machine learning integration, while C++ supports high-performance simulation of cryptographic primitives and protocol execution.

AI models, including transformer-based architectures and generative adversarial networks (GANs), are trained on curated corpora—such as the Penn Treebank, WikiText, and the Cornell ArXiv archive—to generate context-aware adversarial probes and identify semantic patterns associated with cryptographic protocol misuse. These models are embedded into a modular simulation environment where test cases are dynamically generated and assessed against the targeted quantum protocols.

To evaluate protocol resilience, the framework integrates automated fuzzing tools (e.g., AFL, libFuzzer), real-time anomaly detection (e.g., Isolation Forest, PCA), and side-channel simulation modules. Each test iteration is designed to expose latent vulnerabilities and implementation-level weaknesses, with results fed into an iterative feedback loop to enhance protocol robustness. Vulnerabilities uncovered through these methods undergo structured forensic analysis, leading to the design of compensating controls and the refinement of security configurations.

The methodology is grounded in adversarial simulation and continuous validation, with "red teaming" employed not as a singular exercise but as a persistent, adaptive threat modelling cycle. By systematically probing cryptographic implementations under varied attack conditions, the approach ensures a high-fidelity assessment of both protocol-level and system-level defences.

The flowchart in Figure 2 provides a schematic overview of the red teaming methodology, illustrating the flow from initial theoretical modelling to real-time threat simulation and feedback-driven refinement. This visual framework supports reproducibility and offers a structured path for deploying AI-augmented penetration testing techniques in quantum cybersecurity contexts.

**Dr. Petar Radanliev**
Parks Road,
Oxford OX1 3PJ
United Kingdom
Email: petar.radanliev@cs.ox.ac.uk
BA Hons., MSc., Ph.D. Post-Doctorate

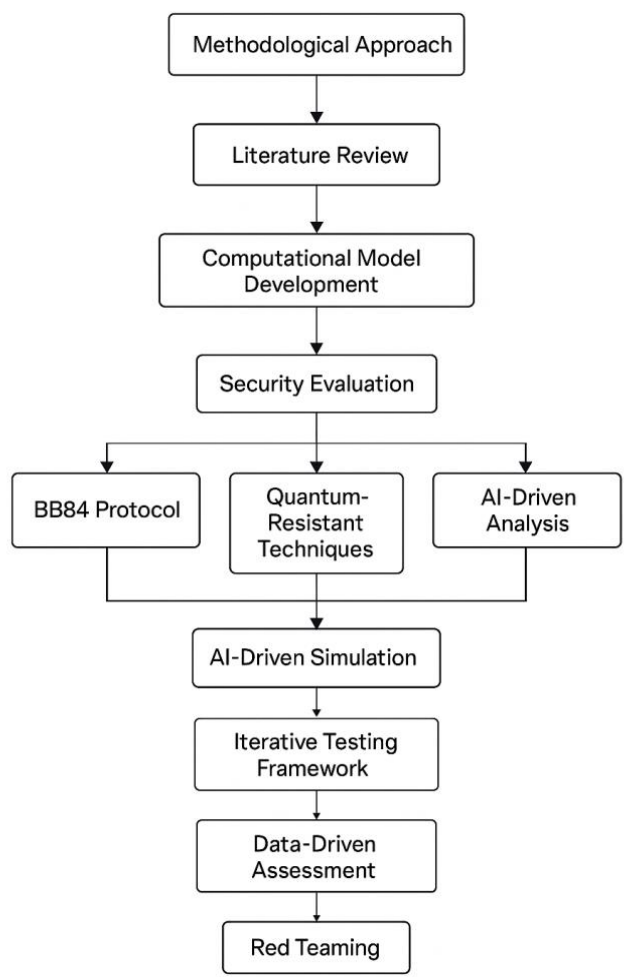

*Figure 2: Methodological framework for AI-driven red teaming of quantum cryptographic protocols.*

The diagram illustrates the structured progression of the research methodology, beginning with a literature review and computational model development, followed by security evaluation across the BB84 protocol, quantum-resistant cryptographic techniques, and AI-driven analysis. These stages inform iterative AI simulations, which feed into a testing framework and data-driven assessment, culminating in red teaming to evaluate and refine protocol resilience.

## 3. The research gap and methodological objectives

Despite significant advancements in cryptographic research, a critical gap persists in the methodological assessment of quantum-resistant algorithms when subjected to AI-driven adversarial conditions. While foundational cryptographic models, including symmetric and asymmetric schemes, are well-understood, their quantum counterparts and post-quantum variants remain insufficiently stress-tested under

Dr. Petar Radanliev
Parks Road,
Oxford OX1 3PJ
United Kingdom
Email: petar.radanliev@cs.ox.ac.uk
BA Hons., MSc., Ph.D. Post-Doctorate

real-world operational scenarios, particularly through dynamic red teaming approaches. The urgency of this challenge is amplified by the emergence of quantum computing, which threatens to render many widely used encryption schemes obsolete. Moreover, current evaluation frameworks largely neglect the integration of automated exploit simulations, adaptive adversarial learning, and protocol fuzzing tools within cryptographic testing pipelines. The objective of this study is to address this gap by developing a structured methodology that incorporates machine learning, quantum-aware simulation, and red teaming to evaluate the resilience of BB84 and NIST-endorsed quantum-safe algorithms. This framework aims to expose latent vulnerabilities and to operationalise AI-assisted refinements that contribute to the development of robust, forward-compatible security protocols.

## 3.1. Cryptography

Within the broader objective of this study, to evaluate and reinforce the resilience of quantum cryptographic protocols using AI-driven red teaming, the role of classical cryptographic primitives remains foundational. Understanding the structural logic, operational dependencies, and failure modes of symmetric and asymmetric encryption schemes is critical to identifying where post-quantum and quantum-enhanced threats are likely to manifest. Traditional cryptographic systems are generally evaluated based on three principal attack surfaces: (1) the mathematical hardness of the algorithm, (2) the correctness and security of the implementation, and (3) the confidentiality and integrity of key management systems. Of these, only the first is inherently cryptographic in nature, while the latter two represent systemic vulnerabilities that can be exploited via adversarial machine learning, protocol fuzzing, and targeted implementation-level attacks, precisely the focus of the red teaming methodology employed in this research.

In classical systems, the cryptographic algorithm itself is rarely the weakest link. Instead, operational misuse, implementation errors, or key leakage tend to expose the system to adversarial compromise. These realities provide the rationale for integrating adversarial simulations into cryptographic assessment, allowing AI agents to systematically explore protocol edge cases, implementation assumptions, and potential misuse patterns under both classical and quantum threat models.

### 3.1.1. Symmetric

Symmetric key encryption schemes, such as the Advanced Encryption Standard (AES), are based on the use of a single shared key for both encryption and decryption. AES (also known as Rijndael [2]), formally standardised by NIST in 2001 [3], has become the global benchmark for secure high-throughput encryption. However, symmetric schemes inherently require a secure mechanism for key exchange, a longstanding bottleneck in distributed systems and a key motivation behind the development of quantum key distribution (QKD). In this study, symmetric encryption is not examined in isolation, but rather in its interaction with QKD protocols (such as BB84), where AI models are trained to simulate both benign and malicious communication behaviours. By using automated anomaly detection and red teaming simulations, the resilience of symmetric ciphers under adversarial quantum environments is empirically tested. The AI agents are tasked with identifying leakage

Dr. Petar Radanliev
Parks Road,
Oxford OX1 3PJ
United Kingdom
Email: petar.radanliev@cs.ox.ac.uk
BA Hons., MSc., Ph.D. Post-Doctorate

vectors during key exchange and testing protocol robustness under noise, interference, or implementation defects.

### 3.1.2. Asymmetric

Asymmetric cryptography is also known as public-key cryptography, uses two different keys, one is public key that is used for encryption and is known to all, and second is the private key that is used for decryption and is only known by one party. The most famous algorithm for public-key cryptography is the RSA cryptosystem developed in 1977 [4] Other well-known and frequently used algorithms include: the Digital Signature Algorithm (DSA), which is based on the Schnorr and ElGamal signature schemes [5]; the Diffie–Hellman key exchange over public channels [6]; or as others have referred to as a method for 'secure communications over insecure channels' [7]; or the Elliptic-curve cryptography (ECC) that is based on algebraic structure of elliptic curves over finite fields. One point that is quite interesting to mention, while the RSA cryptosystem was publicly described the algorithm in 1977, the British mathematician and cryptographer Clifford Cocks, while working for the GCHQ in the year 1973, described an equivalent system in an internal document [8], what this brings to lights is that knowledge discovery is a process that follows a linear pattern. Hence, although we do not know how to develop or even implement new types of cryptography, the knowledge developed is ongoing, and the question is not whether the solutions will be developed, but who will develop the new solutions first. However, these algorithms are particularly vulnerable to quantum attacks due to their reliance on number-theoretic problems, such as integer factorisation and discrete logarithms, which are rendered solvable in polynomial time by Shor's algorithm [9].

The methodological importance of this observation is twofold. First, it validates the relevance of NIST's post-quantum cryptography (PQC) standardisation initiative, which this study aligns with by embedding NIST-approved lattice-, hash-, and code-based schemes into the simulation testbed. Second, it highlights the necessity of adversarial testing against the algorithmic core, the protocol wrappers, APIs, and communication infrastructure in which these schemes are deployed. The red teaming framework developed in this study includes adversarial models that probe for non-cryptanalytic vulnerabilities: side-channel weaknesses, inconsistent key handling, and misconfigured authentication flows. Through these targeted assessments, the study contributes to a more realistic and implementation-aware evaluation of asymmetric cryptographic resilience in quantum-aware environments.

## 3.2. Quantum cryptography

Quantum cryptography is based on the fundamental principles of quantum mechanics, such as superposition, entanglement, and measurement disturbance, to construct protocols that are theoretically immune to certain classes of computational attacks. Unlike classical cryptographic systems, whose security depends on the intractability of mathematical problems (e.g., factorisation or discrete logarithms), quantum cryptographic protocols derive their security from physical laws. A key feature is that the act of observing a quantum system inevitably alters its state, making undetected eavesdropping theoretically impossible under ideal conditions.

Dr. Petar Radanliev
Parks Road,
Oxford OX1 3PJ
United Kingdom
Email: petar.radanliev@cs.ox.ac.uk
BA Hons., MSc., Ph.D. Post-Doctorate

Central to this framework is the concept of quantum superposition, wherein quantum bits (qubits) can exist simultaneously in multiple states—$|0\rangle$, $|1\rangle$, or any linear combination thereof. This quantum parallelism allows a quantum processor to evaluate many potential solutions in a single computational cycle. For instance, a two-qubit system can process all four basis states ($|00\rangle, |01\rangle, |10\rangle, |11\rangle$) simultaneously. Such capabilities vastly outperform classical systems in certain problem domains and, crucially, reduce the computational cost of breaking many widely used public-key encryption schemes.

However, quantum cryptography is not solely defined by computational speed. Its most critical contribution lies in QKD protocols that enable two parties to establish a shared, secret key with provable detection of eavesdropping. The BB84 protocol, developed by Bennett and Brassard in 1984, remains the foundational QKD scheme [10]. It encodes information in the polarisation states of photons, leveraging the no-cloning theorem and measurement disturbance to ensure that any interception attempt introduces detectable anomalies. While idealised BB84 offers theoretical security, practical deployments must account for photon losses, imperfect detectors, and side-channel vulnerabilities. To address these limitations, variants such as the decoy-state BB84 protocol have been introduced, which use randomly varied signal intensities to mitigate photon number splitting attacks in real-world environments [11].

In the context of this research, BB84 is selected as the primary QKD protocol under evaluation, not only due to its foundational role but also because it allows for systematic penetration testing via simulation. By integrating AI-driven red teaming agents, we simulate adversarial behaviours, such as probing for implementation flaws or side-channel emissions, to empirically evaluate BB84's robustness under noisy and compromised conditions. These simulations are conducted within a modular testbed designed to emulate physical-layer quantum channels, enabling detailed anomaly detection and protocol fuzzing strategies.

An increasingly relevant challenge in the quantum security setting is the vulnerability of resource-constrained devices, particularly those in the Internet of Things (IoT) ecosystem. Many embedded systems lack the computational overhead necessary to implement standard cryptographic primitives securely, let alone quantum-safe alternatives. Recognising this, the NIST launched a global standardisation effort to identify lightweight post-quantum cryptographic algorithms suitable for such constrained environments [12]. The integration of these NIST-endorsed post-quantum algorithms into our testing framework enables comparative assessment against BB84-based QKD under identical adversarial conditions.

### 3.3. Low memory cryptography

IoT systems, which operate with minimal memory, limited computational capacity, and low energy budgets, cannot feasibly implement conventional cryptographic protocols, let alone computationally intensive quantum-safe algorithms. This reality has made lightweight cryptography an urgent area of research, particularly as such devices increasingly mediate critical infrastructure, industrial automation, and real-time sensor networks.

Dr. Petar Radanliev
Parks Road,
Oxford OX1 3PJ
United Kingdom
Email: petar.radanliev@cs.ox.ac.uk
BA Hons., MSc., Ph.D. Post-Doctorate

In response to this challenge, the NIST initiated a global call in 2018 for the development of cryptographic algorithms optimised for low-resource environments. The original request for submission[1] for the NIST lightweight cryptography standard resulted in 57 solutions submitted for review by NIST. After rigorous multi-year analysis, Ascon was selected as the official standard for lightweight cryptography[2], balancing the need for high performance on constrained platforms with strong resistance to known cryptanalytic attacks. While the final standardisation and implementation guidelines are still being refined, Ascon represents a foundational shift towards cryptographic primitives that are both resource-aware and quantum-resilient, at least in terms of design assumptions.

From a systems security perspective, the selection of Ascon is notable for its technical efficiency, and for what it implies about the wider cryptographic ecosystem. NIST's review concluded that *'most of the finalists exhibited performance advantages over legacy standards on various platforms without introducing security concerns'*[3]. This finding is significant when considering that many other standardisation bodies, such as ISO and ENISA, have yet to initiate comparable lightweight cryptography evaluations. As a result, global reliance on NIST's benchmarking and standardisation process is likely to increase, amplifying the need for independent validation, adversarial testing, and cross-jurisdictional cryptographic assurance.

The importance of lightweight cryptographic primitives is particularly acute in devices such as Radio Frequency Identification (RFID) tags, embedded sensors, keyless entry fobs, and micromachines, where computational resources and power availability are strictly limited. These devices form the substrate of modern IoT ecosystems but remain highly exposed to adversarial manipulation due to their inability to support classical cryptographic defences. Compounding this risk is the increasing likelihood that quantum-capable adversaries may target these nodes as soft entry points into broader systems.

In this study, the red teaming methodology is extended to evaluate the operational resilience of Ascon and other lightweight candidates under simulated adversarial conditions. By deploying AI agents trained in side-channel modelling, input fuzzing, and low-level protocol probing, we assess how lightweight cryptographic implementations respond to targeted attack patterns within constrained environments. This includes examining fault injection scenarios, memory overflow edge cases, and timing-based anomaly detection failures. The insights generated through these exercises are used to identify latent vulnerabilities, particularly those that might emerge only in specific hardware-software integration contexts.

---

[1] https://www.nist.gov/news-events/news/2018/04/nist-issues-first-call-lightweight-cryptography-protect-small-electronics

[2] https://www.nist.gov/news-events/news/2023/02/nist-selects-lightweight-cryptography-algorithms-protect-small-devices

[3] https://csrc.nist.gov/News/2023/lightweight-cryptography-nist-selects-ascon

Dr. Petar Radanliev
Parks Road,
Oxford OX1 3PJ
United Kingdom
Email: petar.radanliev@cs.ox.ac.uk
BA Hons., MSc., Ph.D. Post-Doctorate

## 3.4. Risk Management

A central objective of this research is to develop a rigorous framework for adversarial evaluation of quantum-resilient cryptographic protocols. Building on the foundational discussions of BB84-based QKD, NIST-standardised post-quantum algorithms, and lightweight cryptographic primitives for constrained environments, this section advances the study's methodological commitment to risk-oriented red teaming. The approach is not limited to theoretical cryptographic soundness; rather, it focuses on operational vulnerabilities and threat scenarios that emerge in hybrid environments where quantum and AI-driven systems coexist. The integration of AI agents trained to simulate real-world and future-state adversarial behaviours enables the identification, classification, and mitigation of cryptographic failure modes before they manifest in deployed systems.

### 3.4.1. Risk identification

The red teaming methodology developed in this study is designed to evaluate the resilience of quantum and post-quantum cryptographic systems, particularly the BB84 QKD protocol and the NIST-selected algorithms for post-quantum cryptography, against adversarial strategies driven by generative AI and natural language processing models. These AI agents emulate dynamic attack strategies, allowing for the testing of implementation boundaries, key exchange errors, and protocol misconfigurations. The objective is to identify latent weaknesses that may not be apparent in formal proofs but can surface under real-time conditions, especially when deployed in insecure or adversarial operational environments. In high-assurance systems, such as military-grade quantum networks, the early identification of such vulnerabilities is not optional but essential [13].

This adversarial approach enhances the fidelity of cryptographic risk assessment by embedding attack simulations within the system's execution environment, probing for edge cases through techniques such as automated fuzzing, reinforcement learning-based strategy adaptation, and AI-assisted anomaly detection. The outcome is a feedback-driven process that detects vulnerabilities and informs the development of countermeasures, threat models, and quantum-secure design practices.

This, in conjunction with quantum cryptographic procedures, will predominantly focus on the BB84 protocol [14] and the suite of NIST-endorsed Quantum-Resistant Cryptographic Algorithms [15,16]. The study outlines the penetration testing phases and the salient objectives that encompass identifying and rectifying frailties within the BB84 protocol, thereby refining its cryptographic resilience and culminating in developing a fortified quantum-secure prototype. The integration of quantum cryptographic techniques into military systems has already been explored as a means to shield software against quantum-powered cyberattacks, underscoring the importance of adopting advanced cryptographic methods in defence contexts [13].

### 3.4.2. Threat Scenarios: Quantum vs. Post-Quantum Risk Vectors

Two distinct but interrelated risk domains are central to the methodological scope of this research: (1) risks associated with future large-scale quantum computers, and (2) risks arising from retrospective exploitation of current cryptographic systems, both of which threaten to undermine today's digital infrastructure.

Dr. Petar Radanliev
Parks Road,
Oxford OX1 3PJ
United Kingdom
Email: petar.radanliev@cs.ox.ac.uk
BA Hons., MSc., Ph.D. Post-Doctorate

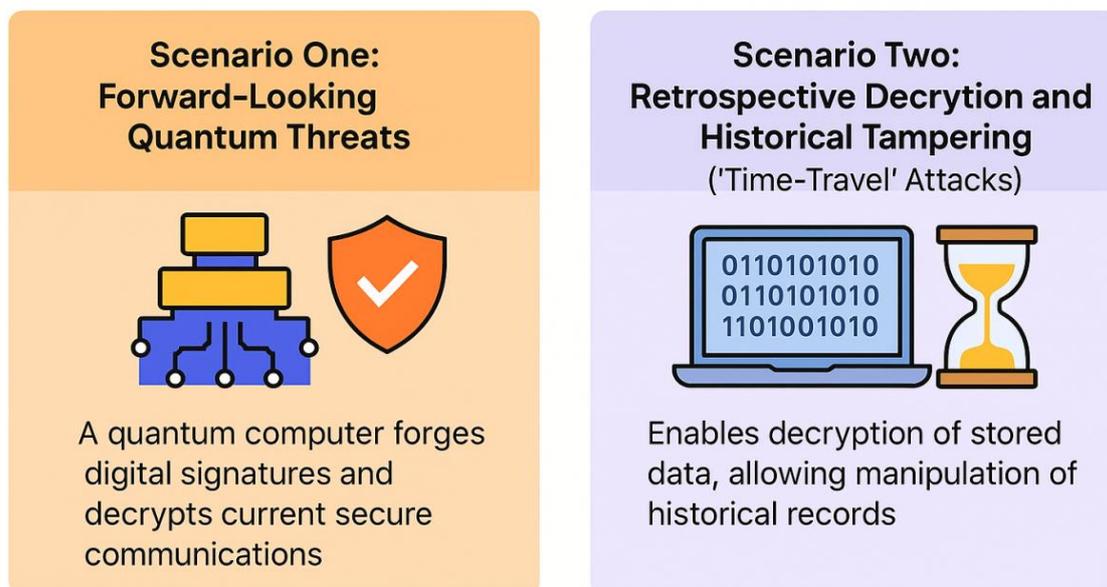

*Figure 3: Dual Risk Scenarios of Quantum Cryptographic Threats*

Figure 3 visualises two critical scenarios posed by the emergence of quantum computing. Scenario One represents forward-looking quantum threats, wherein a large-scale quantum computer could undermine contemporary public-key cryptographic systems such as RSA, ECC, and blockchain-based protocols. Scenario Two highlights the risk of retrospective decryption attacks, often termed 'time-travel' threats, where encrypted data intercepted and stored today could be decrypted in the future, compromising the historical integrity of medical, financial, and governance records. Together, these scenarios underscore the need for quantum-resilient cryptographic standards and proactive mitigation strategies.

**Scenario One: Forward-Looking Quantum Threats**

In this scenario, a future quantum computer, equipped with sufficient qubit coherence and fault tolerance, would be capable of breaking the majority of public-key cryptographic schemes in use today [9], including RSA, ECC, and protocols used in blockchain systems (e.g., EC/EdDSA, VRFs, ZK proofs). With such capabilities, an adversary could forge digital signatures, extract private keys, or decrypt secure communications, leading to catastrophic failures in digital finance, e-government, and industrial control systems. While this remains a theoretical risk for now, its inevitability has already motivated NIST to initiate the PQC standardisation process in 2016.

**Scenario Two: Retrospective Decryption and Historical Tampering ('Time-Travel' Attacks)**

Dr. Petar Radanliev
Parks Road,
Oxford OX1 3PJ
United Kingdom
Email: petar.radanliev@cs.ox.ac.uk
BA Hons., MSc., Ph.D. Post-Doctorate

The second and perhaps more urgent risk involves the deferred threat of retrospective quantum attacks. Data encrypted today with classically secure algorithms may be stored by malicious actors and decrypted in the future once a sufficiently powerful quantum computer becomes available. This retrospective decryption would enable historical manipulation of digital records, ranging from medical files and financial transactions to blockchain consensus states, introducing a novel class of long-range integrity and authenticity risks.

*Post-Quantum Mitigation: Verifiable State Anchoring via SNARK-Compatible Proofs.*

Innovative approaches, such as employing deep reinforcement learning to develop authentication and key establishment protocols [17], have been proposed to enhance resilience against quantum computing threats [18].

To mitigate these risks, particularly the second scenario, this study explores the integration of State Proofs as a method for anchoring the state of digital assets in a way that is verifiable, decentralised, and compatible with post-quantum constraints. State Proofs are structured digital attestations that provide cryptographic evidence of the system's state at a given time. They serve as lightweight certificates that can be externally verified without relying on the underlying system's ongoing operational trust.

The architecture of State Proofs proposed in this research is designed to be:

1. Low-cost to verify, enabling validation even on constrained devices;
2. Detached from the main network, allowing out-of-band attestation and forensic analysis;
3. SNARK-friendly, so that proofs can be succinct, composable, and compatible with zero-knowledge infrastructures.

In practical terms, State Proofs would be periodically generated and embedded into a smart contract environment, with each proof referencing the cumulative attested stake or consensus state. To prevent forgery, even by an actor equipped with quantum capabilities, the design assumes that generating a valid proof requires contributions from a supermajority (e.g., 70–80%) of the network. This assumption ensures that even if a future quantum adversary possesses computational superiority, they cannot fabricate a state history without collaboration from the honest majority. Verification is decentralised, and the aggregation of proofs is performed by an untrusted coordinator that holds only partial views, similar in principle to two-factor authentication but resistant to centralised compromise.

Where applicable, the implementation of these proofs relies on deterministic signature schemes like Falcon [15], known for their linear verification complexity and SNARK compatibility. These design choices ensure that State Proofs can be efficiently verified by external agents, contributing to quantum-resistant assurance at the system level.

*The benefits*

The benefits of this solution are proofs of the state that can be implemented in the networking protocols and architecture for easy verification of the state by entities

**Dr. Petar Radanliev**
Parks Road,
Oxford OX1 3PJ
United Kingdom
Email: petar.radanliev@cs.ox.ac.uk
BA Hons., MSc., Ph.D. Post-Doctorate

outside of the network. The solution is based on distributed quantum computing and adds long-term post-quantum security to the networking protocols and architecture. The implementation can be ultra-compressed into tiny and cheap SNARKs. This solution adds long-term post-quantum security because to create a proof of state in a distributed system, you need to have a certain contribution from the network, a fraction (around 70-80%) of the stake attested. Without that fraction, it is infeasible to create a stake proof, even if you had a quantum computer. In other words, if a quantum computer tries to create a fake proof of stake, the 'State Proofs' would confirm the previous state. The solution also improves network interoperability, because by converting the proof of state into a compressed SNARK (e.g., zk-SNARK proofs). The solution enhances protocol transparency, strengthens forensic traceability, and supports network-level interoperability through compressed SNARK proofs. From a systems engineering perspective, this approach bridges the methodological gap between red teaming, runtime verification, and post-quantum readiness, ensuring that cryptographic systems are not only theoretically sound but also resilient under adversarial conditions in both classical and quantum threat landscapes.

## 4. Testing process

Our postulate asserts that a specialised red team approach, merging AI/NLP blueprints with the quantum cryptographic principles of the BB84 protocol and NIST-sanctioned algorithms, can protect quantum internet infrastructure [19]. In accord and conjunction with the White House's Office of Science, Technology, and Policy, we suggest a dedicated forensic assessment of these emergent generative AI constructs. With the White House's explicit endorsement for such autonomous evaluative endeavours [20], we posit our methodology, rooted in Red Teaming paradigms, as a beacon aligning with the foundational principles of the Biden administration's AI Bill of Rights [21] and the AI Risk Management edicts decreed by the National Institute of Standards and Technology [22]. By definition, "red teaming" encapsulates a proactive security, where specialists assume adversarial roles to challenge, evaluate, and enhance the defensive robustness of systems and frameworks.

**Dr. Petar Radanliev**
Parks Road,
Oxford OX1 3PJ
United Kingdom
Email: petar.radanliev@cs.ox.ac.uk
BA Hons., MSc., Ph.D. Post-Doctorate

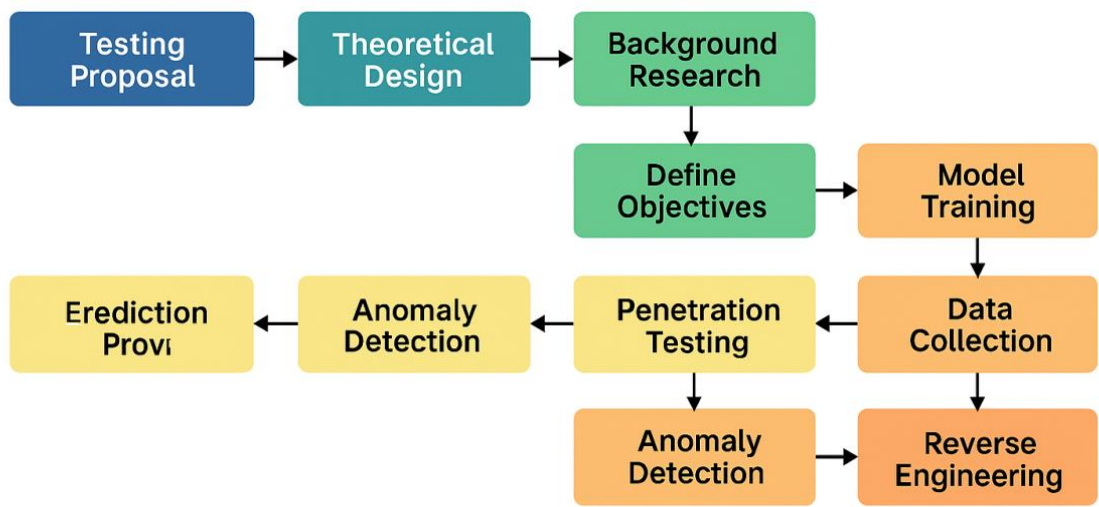

*Figure 4: Penetration Testing Approaches for the BB84 Quantum Cryptography Protocol*

Figure 4 synthesises the multi-layered testing methodologies employed to evaluate the security resilience of the BB84 quantum key distribution protocol. It visually categorises eight advanced approaches, ranging from traditional penetration testing to AI-driven red teaming, quantum-aware exploit simulation, adversarial machine learning, and real-time anomaly detection. Each method is mapped against its key benefits, limitations, technical models, and sequential testing steps. The framework supports a comprehensive evaluation of both classical and quantum-specific vulnerabilities, integrating physical-layer simulations and post-processing validation to align with emerging AI and quantum security standards.

In Table 1, the flowchart provides a visual representation of the research methodology of the testing process, starting with the initial testing proposal and moving through various stages, including theoretical design, background research, objectives definition, model training, environment setup, penetration testing, data collection, anomaly detection, reverse engineering, and feedback integration.

*Table 1: Penetration testing approaches for the BB84 quantum cryptography protocol*

| Approach | Description | Key Benefits | Limitations | Methods & Models | Specific Steps |
|---|---|---|---|---|---|
| **Traditional Penetration Testing** | Conventional security testing methods applied to BB84, involving manual and automated | Establishes baseline security but lacks adaptability to quantum-specific threats. | Not optimised for quantum-specific attacks, may overlook key | Manual testing, automated scanning, penetration testing frameworks (e.g., | 1. Identify BB84 protocol implementation. 2. Perform threat modelling. 3. Execute |

Dr. Petar Radanliev
Parks Road,
Oxford OX1 3PJ
United Kingdom
Email: petar.radanliev@cs.ox.ac.uk
BA Hons., MSc., Ph.D. Post-Doctorate

| | | | | | |
|---|---|---|---|---|---|
| | exploit attempts. | | vulnerabilities. | Metasploit, Kali Linux). | simulated attack scenarios. 4. Analyse protocol responses. 5. Report vulnerabilities. |
| **AI-Driven Red Teaming** | Utilisation of AI models to generate and adapt red teaming strategies against BB84-based cryptographic implementations. | Enhances security testing with AI-driven adaptivity and scalability. | Dependent on AI model quality and training data; risks generating false positives. | Reinforcement learning-based attack simulations, generative adversarial networks (GANs) for adaptive exploits. | 1. Train AI models on attack scenarios. 2. Generate adversarial test cases. 3. Simulate red teaming attacks. 4. Adapt strategy based on system response. 5. Evaluate AI-driven attack efficiency. |
| **Quantum-Aware Exploit Simulation** | Simulation of quantum-specific exploits, targeting weaknesses in the physical and computational aspects of BB84. | Provides insights into BB84's resistance to quantum-specific attack vectors. | Limited by current quantum exploit knowledge and hardware constraints. | Quantum emulator testing, fault injection models, quantum circuit-level vulnerability scanning. | 1. Develop quantum exploit test cases. 2. Implement fault injection in quantum key exchange. 3. Simulate quantum-specific attacks. 4. Assess protocol stability. 5. |

Dr. Petar Radanliev
Parks Road,
Oxford OX1 3PJ
United Kingdom
Email: petar.radanliev@cs.ox.ac.uk
BA Hons., MSc., Ph.D. Post-Doctorate

| | | | | | Identify failure points. |
|---|---|---|---|---|---|
| **Cryptographic Vulnerability Assessment** | Evaluation of potential cryptographic weaknesses in BB84, assessing key exchange integrity and quantum noise resilience. | Ensures cryptographic robustness against classical and quantum adversaries. | May not fully capture implementation-specific weaknesses in BB84 deployments. | Mathematical proof-based security analysis, quantum key distribution (QKD) resilience testing. | 1. Perform formal cryptographic analysis. 2. Test key generation randomness. 3. Simulate noise-induced errors. 4. Assess error correction performance. 5. Validate cryptographic robustness. |
| **Adversarial ML Attacks on BB84** | Testing adversarial AI attacks against AI-enhanced BB84 implementations, probing for vulnerabilities in decision models. | Examines AI's influence on BB84 security, identifying novel attack surfaces. | Requires robust adversarial training datasets for meaningful assessments. | Adversarial neural network training, perturbation analysis on quantum-enhanced AI models. | 1. Select adversarial perturbation models. 2. Apply adversarial noise to BB84 AI components. 3. Monitor changes in protocol behaviour. 4. Quantify attack success rate. 5. Refine AI training models. |

Dr. Petar Radanliev
Parks Road,
Oxford OX1 3PJ
United Kingdom
Email: petar.radanliev@cs.ox.ac.uk
BA Hons., MSc., Ph.D. Post-Doctorate

| | | | | | |
|---|---|---|---|---|---|
| **Automated Protocol Fuzzing** | Systematic probing of BB84 protocol variations using automated input generation to identify edge-case failures. | Automates vulnerability detection, increasing efficiency in protocol testing. | Potentially resource-intensive, requiring high computational overhead. | Fuzzing engines (e.g., AFL, libFuzzer), mutation-based protocol stress testing. | 1. Define BB84 protocol variations. 2. Generate random and structured test cases. 3. Execute fuzzing iterations. 4. Analyse unexpected protocol failures. 5. Patch identified weaknesses. |
| **Side-Channel Attack Simulation** | Analysis of power consumption, electromagnetic leaks, and timing patterns to detect potential side-channel vulnerabilities. | Identifies potential hardware and implementation-level weaknesses. | Highly dependent on hardware setup and precision in data collection. | Differential power analysis, timing attack models, electromagnetic analysis tools. | 1. Collect hardware leakage signals. 2. Apply statistical analysis to detect patterns. 3. Correlate side-channel data with cryptographic operations. 4. Develop mitigation strategies. 5. Implement hardware-level protections. |

Dr. Petar Radanliev
Parks Road,
Oxford OX1 3PJ
United Kingdom
Email: petar.radanliev@cs.ox.ac.uk
BA Hons., MSc., Ph.D. Post-Doctorate

| **Real-Time Anomaly Detection** | Implementation of real-time monitoring and ML-based anomaly detection to identify deviations from expected BB84 protocol behaviours. | Improves detection of active threats through dynamic behavioural analysis. | May introduce latency in security responses due to processing overhead. | Supervised and unsupervised ML models, anomaly detection frameworks (e.g., Isolation Forest, PCA-based detection). | 1. Establish behavioural baselines. 2. Train ML models on normal BB84 operations. 3. Monitor for deviations from expected patterns. 4. Generate alerts on anomalous activity. 5. Automate response mechanisms. |

These technologies listed in Table 1, could be of particular interest to cybersecurity firms and agencies, academic researchers, and any industries where secure, fast communication is essential. One specific comment on the table is related to the assess error correction performance (point 4 in the section 'Cryptographic Vulnerability Assessment'), the penetration testing approaches that we considered for the BB84 quantum cryptography protocol, are based on the fact that a modular simulation framework enables detailed analysis of QKD post-processing stages, including error correction and privacy amplification, allowing for improved modelling of end-to-end cryptographic performance [23]. Second comment that needs to be emphasises in relation to the penetration testing approaches that we considered, is related to the section simulation of quantum-specific exploits, targeting weaknesses in the physical and computational aspects of BB84, which related to the category 'Quantum-Aware Exploit Simulation', and in relation to this category, it needs to be emphasised that modelling quantum optical components has revealed potential implementation-level vulnerabilities in QKD systems, underscoring the importance of physical-layer simulations for comprehensive red teaming strategies [17].

## 4.1. Red Teaming design

As the threat landscape evolves, it is crucial to take proactive measures to identify and address vulnerabilities. AI-driven methods have the potential to redefine cybersecurity standards, making systems more reliable and secure. It is crucial to establish secure protocols that can withstand quantum attacks. By enabling

Dr. Petar Radanliev
Parks Road,
Oxford OX1 3PJ
United Kingdom
Email: petar.radanliev@cs.ox.ac.uk
BA Hons., MSc., Ph.D. Post-Doctorate

enhanced quantum security, we can instil confidence in the confidentiality and safety of quantum communications, leading to greater trust and adoption of this technology.

Automated quantum pen-testing kits have been created to streamline evaluating their security. These advanced kits are engineered to automatically test the security of quantum systems, providing users with an overview of their current security status.

Cutting-edge solutions have emerged to tackle cybersecurity challenges, harnessing the power of AI to optimise reverse engineering tasks and facilitate payload delivery systems that combat quantum exploits. Reverse engineering is crucial for identifying potential system weaknesses, and AI integration allows for greater precision and efficiency in detecting vulnerabilities. Advanced payload delivery systems ensure comprehensive security assessments addressing known and unknown threats.

### 4.2.   Ethical penetration testing

Our primary objective is to establish a strong and reliable framework for the upcoming quantum internet risks. To achieve this, we focus on refining and enhancing the BB84 protocol, in conjunction with NIST-approved algorithms [24]. System-level modelling of decoy-state QKD confirms its effectiveness in enhancing BB84's resistance to eavesdropping, further validating the need for advanced red teaming frameworks in practical quantum networks [25]. We aim to ensure that all data transmissions remain secure and tamper-proof, which is crucial for building trust in digital communication.

### 4.3.   Prototyping & Development

The process of refinement of quantum cryptographic mechanisms, remains on the adaptation and elevation of the BB84 protocol [14] and other NIST-endorsed quantum cryptographic methodologies, algorithms [15,16], and cryptographic mechanisms [15,26,27]. With AI integration, we can expect new algorithms that can be interconnected with quantum frameworks [28].

The simulation of quantum-secured environments using AI/NLP models includes generative AI in simulating conventional and malevolent user behaviours within a quantum network environment [29]. Foundational datasets like the Cornell ArXiv, supplemented by the Penn Treebank and WikiText, will serve as the bedrock for training our models in cryptographic contexts [30]. Our methodology is based on implementing NLP techniques, with a specific emphasis on transformer-based models such as GPT variants [31]. The robustness and versatility of libraries like HuggingFace's Transformers will be key to ensure the efficacy of AI models in BB84 quantum cryptography simulation. Potential integrations include platforms like Qiskit or QuTiP, in the simulation cycles. Programmable network interfaces such as OpenFlow have been shown to effectively manage quantum metadata in QKD networks, offering a scalable solution for adaptive control of quantum communications infrastructure [32]. Python can be used for scripting, automation, and data aggregation of retesting scenarios [33].

**Dr. Petar Radanliev**
Parks Road,
Oxford OX1 3PJ
United Kingdom
Email: petar.radanliev@cs.ox.ac.uk
BA Hons., MSc., Ph.D. Post-Doctorate

The assessment of NIST quantum-resistant cryptographic algorithms includes the integration of Lattice-based cryptographic methods to Code-based encryption techniques and Hash-based signatures.

Quantum computing requires evaluating quantum systems' and protocols' real-world efficacy and vulnerabilities. This research constructs a theoretical framework for this purpose.

**Conceptual Foundations**:

1. **Quantum Network Dynamics**: Drawing from foundational principles of quantum mechanics and network theory, we postulate quantum networks' potential behaviours and challenges in real-world settings.
2. **User Interaction with Quantum Systems**: Grounded in human-computer interaction theories, we explore the nuances of end-user engagement with quantum systems, focusing on usability and potential user-triggered vulnerabilities.

**Data Sources and Methodological Considerations**:

1. **Collaborative Simulations**: By partnering with industry leaders, we aim to simulate authentic network scenarios, bridging the gap between theoretical postulations and practical applications.
2. **Synthetic Data Generation**: This approach, rooted in predictive modelling, seeks to emulate future quantum network behaviours, offering insights into prospective challenges and solutions.

**Proposed Theoretical Constructs**:

1. **AI/NLP-Driven Quantum Network Behaviours**: Integrating AI/NLP models with quantum simulations offers a novel perspective on network traffic behaviours, typical and adversarial.
2. **User-Centric Quantum System Design**: By understanding end-user interactions and feedback, we can theorise optimal designs for quantum systems that are secure and user-friendly.

**Evaluation and Knowledge Development**:

1. **Performance Metrics in Quantum Networks**: We can develop theories on optimal quantum network designs by identifying key indicators such as detection efficacy and system robustness.
2. **User Feedback Analysis**: A qualitative exploration of user feedback will contribute to the theoretical understanding of user needs, challenges, and potential system improvements in the quantum realm.

In Figure 5, we can visualise the emerging theoretical framework for penetration testing Generative AI and Quantum computers.

Dr. Petar Radanliev
Parks Road,
Oxford OX1 3PJ
United Kingdom
Email: petar.radanliev@cs.ox.ac.uk
BA Hons., MSc., Ph.D. Post-Doctorate

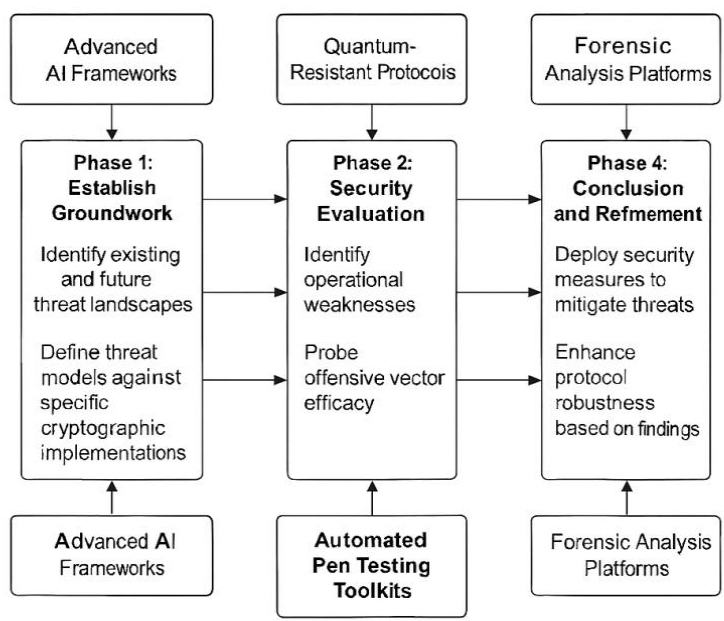

*Figure 5: Framework for penetration testing the merging of Quantum Computing and Artificial Intelligence.*

The framework presented in (in Figure 5) is structured upon theoretical understanding of quantum networks, incorporating iterative refinement, standardisation, and methodical assessment. This design is predicated on the principle that quantum-AI systems must remain adaptable and resilient in the face of evolving computational and security challenges. By drawing upon established methodologies in iterative system design, the framework ensures that continuous development is embedded within the lifecycle of quantum-AI applications. The emphasis on research documentation and standardisation is particularly significant, as systematic recording and transparent reporting are essential to fostering replicability and intellectual integrity in the field. In this context, the framework seeks to establish a structured approach to the evaluation, optimisation, and advancement of quantum-AI architectures.

A core aspect of this framework is its reliance on empirical data and analytical precision. The methodology integrates findings from field trials, user acceptance testing, and emerging research, thereby constructing a comprehensive basis for iterative refinement. The integration of advanced computational tools, such as Python for statistical analysis and C++ for high-performance computing, ensures that the approach remains rigorous and scalable. By employing this dual-layered analytical strategy, the framework is capable of systematically identifying performance bottlenecks and implementing targeted enhancements. The process of optimisation is thereby driven by empirical validation, rather than speculative inference, reinforcing the robustness of the methodology.

The framework advances the notion that the continual optimisation of quantum-AI systems must be grounded in structured performance assessment. The inclusion of real-time feedback mechanisms and benchmarking against established performance

Dr. Petar Radanliev
Parks Road,
Oxford OX1 3PJ
United Kingdom
Email: petar.radanliev@cs.ox.ac.uk
BA Hons., MSc., Ph.D. Post-Doctorate

indicators allows for iterative refinement that is data-driven and methodologically sound. This ensures that quantum-AI models remain at the forefront of computational efficiency while maintaining a strong defence against emerging security threats. A critical component of this refinement process is the structured documentation of quantum research. The collation of research notes, structured datasets, and performance evaluations serves not only to establish transparency but also to provide a robust foundation for future work in this domain. The principle of rigorous documentation ensures that knowledge is preserved, interrogated, and built upon, rather than being lost to ad hoc development cycles.

The framework extends beyond system refinement and addresses the broader question of quantum security through a structured approach to red teaming. In high-assurance computing environments, adversarial testing is indispensable to ensuring system robustness. The approach adopted here is underpinned by an engagement with established security methodologies, incorporating the perspectives of key stakeholders, security analysts, and domain experts. This multi-perspective approach strengthens the integrity of red teaming exercises, ensuring that security assessments are comprehensive and reflective of real-world risks. The incorporation of adaptive threat modelling, in which adversarial simulations evolve in response to defensive countermeasures, ensures that security testing is not confined to static attack scenarios but instead reflects the fluid nature of computational threats.

A key aspect of this security approach is the iterative refinement of countermeasures. By systematically identifying vulnerabilities and incorporating continuous feedback, security mechanisms can be enhanced in a structured and methodical manner. This approach avoids the pitfalls of reactive security, instead fostering a system in which defences are continuously strengthened in response to identified risks. Furthermore, the engagement with ensemble learning techniques allows for the synthesis of insights from multiple analytical models and expert evaluations, ensuring that threat simulations are informed by a breadth of expertise. This approach facilitates a richer and more comprehensive understanding of potential attack vectors, reinforcing the resilience of quantum-AI systems.

The integration of structured evaluation metrics within the framework ensures that security improvements and computational refinements are validated through rigorous assessment. Comparative performance analyses allow for an objective appraisal of iterative improvements, ensuring that refinements are substantiated by empirical evidence rather than theoretical conjecture. The role of peer review in this process is equally significant, as external validation provides an additional layer of scrutiny, ensuring that methodological transparency and reproducibility are maintained throughout the development cycle.

By synthesising principles of iterative refinement, methodological standardisation, empirical validation, and adversarial testing, this framework establishes a robust foundation for the advancement of quantum-AI systems. The structured approach to continuous optimisation and security enhancement ensures that these systems remain computationally efficient and resilient against emerging threats. In doing so, this work contributes not only to the theoretical development of quantum-AI security but also to its practical implementation in high-assurance computing environments.

Dr. Petar Radanliev
Parks Road,
Oxford OX1 3PJ
United Kingdom
Email: petar.radanliev@cs.ox.ac.uk
BA Hons., MSc., Ph.D. Post-Doctorate

This approach, rooted in methodical inquiry and empirical validation, provides a model for ensuring the long-term reliability and security of quantum-AI architectures.

### 4.4. Framework for Quantum Network Behaviour Simulation and Refinement

The progression of quantum computing requires examination of user interactions within quantum networks, particularly in identifying and mitigating malicious behaviours. As quantum architectures become more prevalent, the need for robust analytical frameworks to simulate and interpret these interactions becomes increasingly urgent. This study proposes a theoretical framework designed to structure the modelling and analysis of quantum network behaviour, with a specific emphasis on the BB84 protocol and NIST-endorsed quantum-resistant cryptographic algorithms. By integrating principles from quantum mechanics, network theory, and computational security, this framework establishes a structured approach to understanding the interplay between user behaviours, cryptographic protocols, and potential vulnerabilities within quantum environments.

A central focus of this framework is the delineation of quantum network behaviour dynamics. Drawing from established theories of user interaction and computational modelling, the framework examines the ways in which users engage with quantum systems, distinguishing legitimate actions from potentially adversarial activities. By incorporating methodologies from behavioural analysis and anomaly detection, the framework provides a foundation for identifying deviations from expected usage patterns, thereby enabling a systematic assessment of security risks. In this context, AI and NLP models refine quantum cryptographic protocols, and quantum algorithms for logic-based sampling can be applied to generate representative adversarial scenarios, enabling probabilistic reasoning over quantum-enhanced security models [34]. These models are conditioned to interpret and predict adversarial strategies, allowing for the pre-emptive identification of vulnerabilities in quantum encryption schemes. By employing principles from machine learning, the framework enhances the predictive accuracy of AI-driven security models, strengthening their capacity to detect and respond to emerging threats.

The construction of an emulated quantum environment forms another fundamental aspect of this research. The use of DEVS-based modelling techniques provides a formalised structure for simulating individual QKD components, allowing for modular assessment and iterative refinement of secure quantum systems [35]. The ability to simulate real-world implementations of quantum protocols and cryptographic systems is essential for evaluating their resilience under diverse threat conditions. Discrete event simulation of quantum channels provides an accurate means of modelling time-sensitive behaviours in QKD systems, supporting the evaluation of protocol robustness under dynamic operational conditions [36]. By using established simulation methodologies, the framework enables a controlled environment in which AI/NLP models can be trained to assess the security properties of quantum systems. This approach facilitates the development of adaptive defence mechanisms, ensuring that cryptographic protocols remain robust against evolving attack vectors. The fidelity of these simulations is critical, as they must accurately replicate the

**Dr. Petar Radanliev**
Parks Road,
Oxford OX1 3PJ
United Kingdom
Email: petar.radanliev@cs.ox.ac.uk
BA Hons., MSc., Ph.D. Post-Doctorate

conditions under which quantum networks operate, incorporating theoretical constructs and empirical data to ensure meaningful analysis.

The methodological integrity of this framework is reinforced through active collaboration with stakeholders from academia, industry, and government. Ensuring alignment with real-world quantum security challenges requires continuous engagement with experts who contribute practical insights into the evolving landscape of quantum computing. The framework proposes a structured approach to stakeholder involvement, ensuring that theoretical constructs are rigorously tested against applied security concerns. This collaboration not only enhances the relevance of the research but also fosters the integration of quantum cryptographic advancements into operational environments.

To facilitate AI/NLP model training, this framework incorporates foundational datasets such as Cornell ArXiv, Penn Treebank, and WikiText, which provide a rich repository of structured linguistic and technical data. These resources are instrumental in constructing simulation scenarios that reflect the complexities of quantum cryptographic interactions. By embedding these datasets within the training architecture, AI models gain the ability to discern nuanced security challenges, refining their capacity to detect, analyse, and mitigate potential threats. The integration of structured knowledge sources ensures that AI-driven assessments are informed by a diverse range of empirical and theoretical insights, reinforcing their analytical depth and reliability.

By synthesising behavioural analysis, AI-driven cryptographic modelling, and quantum system simulation, this framework establishes foundation for advancing the study of quantum network security. Figure 6 shows a sequence diagram that includes steps, milestones, loops, critical points, and key outcomes.

Dr. Petar Radanliev
Parks Road,
Oxford OX1 3PJ
United Kingdom
Email: petar.radanliev@cs.ox.ac.uk
BA Hons., MSc., Ph.D. Post-Doctorate

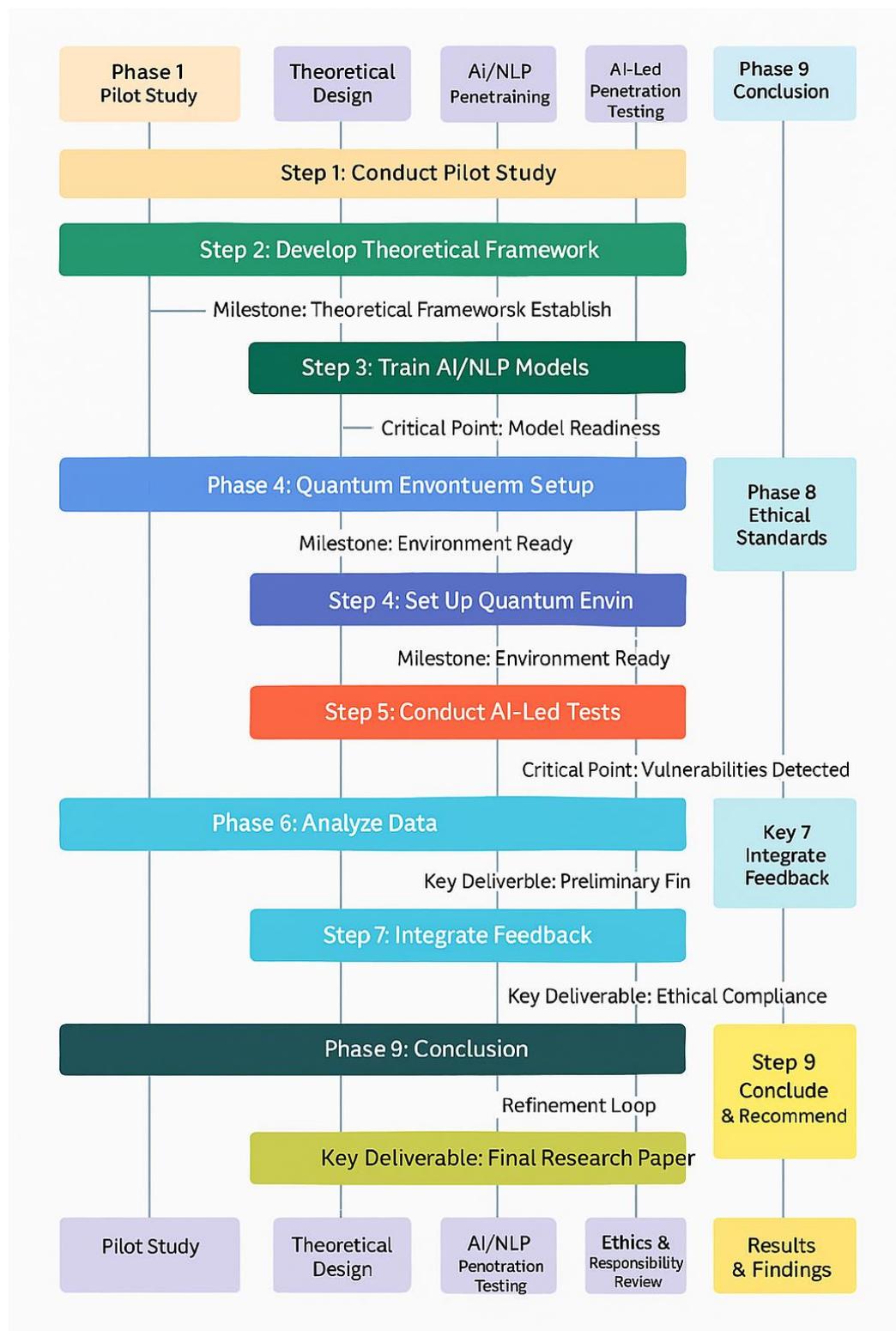

*Figure 6: Sequence diagram outlining the structured methodology for red teaming Generative AI and Quantum cryptographic systems.*

Dr. Petar Radanliev
Parks Road,
Oxford OX1 3PJ
United Kingdom
Email: petar.radanliev@cs.ox.ac.uk
BA Hons., MSc., Ph.D. Post-Doctorate

Figure 6 presents the stepwise progression from pilot study to final recommendations, including critical phases such as AI/NLP model training, quantum environment simulation, AI-led penetration testing, and iterative feedback loops. Distinct colour coding highlights methodological stages, milestones, critical decision points, and deliverables, reflecting the research's dynamic and adaptive workflow aligned with rigorous academic and security standards.

The sequence diagram in Figure 6 provides a structured representation of the methodological process underlying the development and refinement of quantum security mechanisms. This framework is designed to ensure that each stage, ranging from initial environment scanning to iterative optimisation, is theoretically grounded and practically verifiable. The research integrates environment validation, quantum simulation efficiency, and continuous system refinement, establishing a systematic approach for assessing vulnerabilities and reinforcing security in quantum networks.

A key element of this framework is the validation of quantum environments through automated scanning techniques. Python's scripting capabilities are used to conduct real-time assessments, ensuring the isolation and integrity of the quantum infrastructure. The validation process, with C++ providing computational efficiency and Python enabling flexible control structures for scenario testing. Simulation of QKD in fibre-based quantum networks confirms that realistic implementation constraints (such as attenuation and noise) can be effectively modelled to assess protocol resilience under operational conditions [37]. Through iterative refinement, informed by structured feedback loops, the framework systematically adapts to newly identified threats, reinforcing the resilience of quantum networks against adversarial exploits.

The research also advances the development of targeted payloads for assessing quantum system vulnerabilities, integrating principles from cybersecurity and penetration testing. The deployment of these payloads is informed by dynamic analysis techniques, where Python facilitates automation and data handling, while C++ offers precision in execution and low-level control over exploit mechanisms. This dual-layered approach ensures that testing strategies are adaptable and computationally rigorous. Data extracted from these interactions is analysed through statistical and machine learning techniques, offering insights into system weaknesses and enabling the design of effective countermeasures.

Beyond individual exploit testing, the framework incorporates real-time monitoring and anomaly detection mechanisms, ensuring that deviations from expected quantum system behaviours are rapidly identified and assessed. This monitoring process draws upon established theories in anomaly detection and quantum mechanics, enabling the differentiation between benign operational fluctuations and indicators of adversarial activity. To deepen this analytical capacity, the research employs forensic methods for dissecting successful exploits, with reverse engineering tools such as IDA Pro and Ghidra facilitating a granular examination of attack vectors. The insights derived from these assessments inform the development of defensive strategies tailored to the evolving quantum threat landscape.

The role of real-time intelligence in red teaming is further reinforced through data visualisation and interactive reporting methodologies. The framework employs

Dr. Petar Radanliev
Parks Road,
Oxford OX1 3PJ
United Kingdom
Email: petar.radanliev@cs.ox.ac.uk
BA Hons., MSc., Ph.D. Post-Doctorate

Python-based visualisation libraries to present dynamic system metrics, ensuring that red teaming activities are interpreted with clarity and precision. By integrating these reporting techniques into the research workflow, security assessments become more actionable, providing immediate feedback on system vulnerabilities and the effectiveness of implemented countermeasures.

Recognising the iterative nature of quantum security testing, the framework is designed to integrate continuous feedback into its refinement process. Post-testing review mechanisms ensure that findings are systematically analysed and incorporated into subsequent testing cycles. This iterative feedback process is operationalised through structured data pipelines, enabling rapid updates to simulation parameters and security protocols. The efficiency of this approach is enhanced by the computational performance of C++, which facilitates the seamless integration of newly acquired security intelligence into quantum system defences.

The refinement of red teaming methodologies within this framework is underpinned by principles of adaptive security. The synthesis of computational analysis, real-time monitoring, and interactive reporting enables a structured approach. Interactive reporting tools like Jupyter Notebooks for an interactive report [38], allow us to weave together narratives, code snippets, and visuals to inform the review process.

# 5. Discussion: Advancing Quantum-AI Security through Penetration Testing and Red Teaming

The resilience of cryptographic protocols, particularly the BB84 quantum key distribution (QKD) method and NIST-endorsed quantum-resistant algorithms, must be evaluated against emerging quantum cyber threats. The research presented here has developed a penetration testing framework that integrates AI-driven red teaming, automated exploit simulations, and real-time anomaly detection. This framework enhances the assessment of cryptographic vulnerabilities and establishes a structured approach for reinforcing quantum security mechanisms. Performance modelling of space-based QKD protocols reveals specific challenges related to photon loss and noise, which must be accounted for in quantum security frameworks intended for orbital or long-distance communications [39].

The integration of AI in penetration testing introduces a paradigm shift in cybersecurity. Unlike traditional penetration testing methods, which rely on predefined attack scenarios, AI-driven red teaming enables dynamic, adaptive strategies that evolve in response to system defences. By training reinforcement learning models and generative adversarial networks (GANs) to probe BB84 implementations, this research has demonstrated how AI can autonomously identify weaknesses in quantum cryptographic protocols.

A key aspect of this study is the methodological rigor in simulating real-world attack scenarios. The quantum-aware exploit simulation methodology offers an avenue for systematically evaluating protocol vulnerabilities at the intersection of AI and quantum security. By employing quantum emulators, fault injection models, and automated fuzzing techniques, we have identified critical failure points within the BB84 protocol. These findings emphasise the necessity of continuously refining

Dr. Petar Radanliev
Parks Road,
Oxford OX1 3PJ
United Kingdom
Email: petar.radanliev@cs.ox.ac.uk
BA Hons., MSc., Ph.D. Post-Doctorate

quantum security measures to address previously unconsidered vulnerabilities. Furthermore, cryptographic vulnerability assessments provide an additional layer of scrutiny, ensuring that the mathematical foundations of quantum security protocols remain impervious to classical and quantum attacks.

The role of adversarial machine learning in quantum security testing cannot be understated. AI-enhanced cryptographic implementations introduce novel security challenges, particularly when adversarial perturbations are applied to AI-driven key management processes. Through adversarial ML attacks on BB84, this research has explored the potential for AI-driven cryptanalysis to subvert quantum security guarantees. While the BB84 protocol is theoretically secure under ideal conditions, real-world implementations introduce nuances that adversarial AI models can exploit. These findings underscore the need for continuous testing and refinement to maintain the integrity of quantum cryptographic deployments.

Beyond theoretical assessments, this research advances the practical application of security testing methodologies through automated protocol fuzzing and side-channel attack simulations. The automated penetration testing framework, integrating tools such as AFL, libFuzzer, and Python-based anomaly detection libraries, allows for systematic probing of quantum protocol variations. This automation enhances efficiency in security assessments while uncovering vulnerabilities that might otherwise go unnoticed in manual testing scenarios. Additionally, side-channel attack simulations have highlighted the significance of implementation-level security. Even theoretically secure quantum protocols can be compromised through indirect means, such as electromagnetic emissions, timing leaks, and power consumption analysis. This research demonstrates how these attack vectors can be systematically evaluated and mitigated.

Real-time anomaly detection is an essential component of this penetration testing framework. By using supervised and unsupervised machine learning models, deviations from expected BB84 protocol behaviours can be identified with high accuracy. The implementation of anomaly detection frameworks, including Isolation Forest and PCA-based detection methods, allows for early threat identification, reducing the window of opportunity for adversarial exploitation. This approach represents a shift towards proactive cybersecurity, where potential threats are detected and addressed before they can materialise into security breaches.

The iterative refinement of security strategies is another cornerstone of this research. Post-testing review mechanisms ensure that the insights derived from penetration testing exercises translate into actionable improvements in quantum security frameworks. The rapid implementation of feedback loops, facilitated by C++-driven security updates, ensures that quantum systems remain resilient against evolving attack methodologies. This iterative approach aligns with the principles of adaptive security, where continuous threat assessment and response mechanisms are embedded into the development lifecycle of quantum cryptographic protocols.

The broader implications of this research extend beyond the technical refinement of penetration testing methodologies. The structured evaluation of BB84 security not only contributes to the academic discourse on quantum cryptography but also has direct applications in industry, government, and national security contexts. The ability

Dr. Petar Radanliev
Parks Road,
Oxford OX1 3PJ
United Kingdom
Email: petar.radanliev@cs.ox.ac.uk
BA Hons., MSc., Ph.D. Post-Doctorate

to systematically assess and enhance the security of quantum communication networks is critical as organisations prepare for the post-quantum era. This research provides a blueprint for integrating AI-driven cybersecurity strategies into national security infrastructures, ensuring that quantum technologies are deployed in a manner that is secure and resilient.

Furthermore, this study highlights the ethical dimensions of quantum-AI security. As AI-driven penetration testing methodologies become more sophisticated, there is a growing need for ethical considerations in the deployment of these technologies. Ensuring that AI-driven exploit detection and red teaming exercises adhere to ethical standards is paramount, particularly when assessing critical infrastructure security. The responsible development of AI-enhanced security testing tools must be a guiding principle to prevent the misuse of these capabilities.

## 6. Conclusion

This study has developed a rigorous framework for assessing and strengthening the security of quantum cryptographic protocols, with particular emphasis on the BB84 quantum key distribution method and NIST-endorsed quantum-resistant algorithms. By integrating AI-driven red teaming, automated penetration testing, and anomaly detection, this research offers a systematic approach to identifying and mitigating vulnerabilities in quantum networks.

A principal contribution of this work is the demonstration that AI can be used to enhance security assessments beyond traditional, static testing methodologies. The application of machine learning techniques, including adversarial modelling and automated exploit detection, has shown that quantum cryptographic defences must evolve dynamically to counter emerging threats. The use of AI to probe weaknesses in BB84 implementations provides new insights into potential attack vectors, reinforcing the need for continuous refinement of quantum security mechanisms.

The study also highlights the role of automation in scaling security assessments, with automated protocol fuzzing and quantum-aware exploit simulations proving essential for uncovering cryptographic weaknesses efficiently. The iterative refinement of countermeasures, informed by empirical testing and real-world attack simulations, ensures that quantum security frameworks remain resilient against classical and quantum adversaries. These findings underscore the importance of structured and adaptive security measures as quantum computing advances.

Beyond the technical contributions, this research has wider implications for industry, national security, and regulatory frameworks. As quantum computing capabilities expand, organisations must adopt proactive security measures, embedding AI-driven risk assessments into their cybersecurity strategies. The methodologies outlined in this study provide a foundation for such integration, offering a model for pre-emptive security evaluation in high-assurance environments.

Ensuring the long-term security of quantum networks will require continued refinement of these approaches. Future work should explore the integration of reinforcement learning for adaptive defence mechanisms and the development of hybrid AI-quantum security models. As quantum computing reshapes digital

Dr. Petar Radanliev
Parks Road,
Oxford OX1 3PJ
United Kingdom
Email: petar.radanliev@cs.ox.ac.uk
BA Hons., MSc., Ph.D. Post-Doctorate

infrastructure, maintaining a rigorous and evolving security posture will be essential to safeguarding critical systems. This research provides a structured methodology for achieving that objective, reinforcing the necessity of continuous assessment, adaptive security strategies, and interdisciplinary collaboration.

Dr. Petar Radanliev
Parks Road,
Oxford OX1 3PJ
United Kingdom
Email: petar.radanliev@cs.ox.ac.uk
BA Hons., MSc., Ph.D. Post-Doctorate

Dr. Petar Radanliev
Parks Road,
Oxford OX1 3PJ
United Kingdom
Email: petar.radanliev@cs.ox.ac.uk
BA Hons., MSc., Ph.D. Post-Doctorate

**Dr. Petar Radanliev**
Parks Road,
Oxford OX1 3PJ
United Kingdom
Email: petar.radanliev@cs.ox.ac.uk
BA Hons., MSc., Ph.D. Post-Doctorate

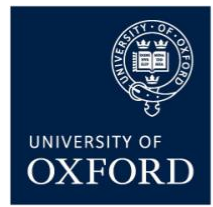

39. Denton JC, Hodson DD, Cobb RG, et al. A model to estimate performance of space-based quantum communication protocols including quantum key distribution systems. *The Journal of Defense Modeling and Simulation* 2019; 16: 5–13.